%% file: Main.tex
\documentclass[preprint,12pt]{elsarticle}

\usepackage[left=3.2cm,right=3.2cm]{geometry}
\usepackage[utf8]{inputenc}
\usepackage{lineno,hyperref}
\usepackage{subcaption}
\usepackage{graphicx}
\usepackage{siunitx}
\usepackage{numprint}
\usepackage[]{xcolor}
\usepackage{amsmath}
\usepackage{amsthm}
\usepackage{amssymb}
\usepackage{mathtools}
\usepackage{dirtytalk}
\usepackage{algorithm2e}
\usepackage{enumitem}
\usepackage{booktabs}
\usepackage{multirow}

\newcommand\norm[1]{\left\lVert#1\right\rVert}
\newcommand{\interval}[2]{[\,#1,#2\,]}
\newcommand{\intsm}[1]{\left[ #1 \right]}
\newcommand{\m}{\text{mid\,}}
\newcommand{\w}{\text{wid\,}}
\newcommand{\bW}{\mathbf{W}}
\newcommand{\bb}{\mathbf{b}}
\newcommand{\bS}{\mathbf{S}}
\newcommand{\bD}{\mathbf{D}}
\newcommand{\by}{\mathbf{y}}
\newcommand{\bX}{\mathbf{X}}
\newcommand{\dw}{\,d\hat{\mathbf{W}}}
\newcommand{\db}{\,d\hat{\mathbf{b}}}

\newcommand{\dr}{\mathrm{d}}
\DeclareMathOperator*{\argmin}{argmin} 

\definecolor{darkgreen}{rgb}{1.0, 1.0, 1.0}
\definecolor{darkred}{rgb}{1.0, 1.0, 1.0}
\newcommand{\reva}[1]{\textcolor{black}{#1}}
\newcommand{\revb}[1]{\textcolor{black}{#1}}

\makeatletter
\renewcommand*\env@matrix[1][\arraystretch]{%
  \edef\arraystretch{#1}%
  \hskip -\arraycolsep
  \let\@ifnextchar\new@ifnextchar
  \array{*\c@MaxMatrixCols c}}
\makeatother

\modulolinenumbers[5]

\journal{}

\begin{document}

\begin{frontmatter}

\title{Neural network model for imprecise regression\\with interval dependent variables
\tnoteref{mytitlenote}}
\tnotetext[mytitlenote]{This work has been partially funded by the Engineering and Physical Science Research Council (EPSRC) through programme grant “Digital twins for improved dynamic design”, EP/R006768/1.}

\author[Liverpool]{Krasymyr Tretiak\corref{cor}}
\ead{k.tretiak@liverpool.ac.uk}

\author[Munich]{Georg Schollmeyer}
\ead{georg.schollmeyer@stat.uni-muenchen.de}

\author[Liverpool]{Scott Ferson}
\ead{ferson@liverpool.ac.uk}

\address[Liverpool]{University of Liverpool,  Liverpool L69 7ZX, United Kingdom}
\address[Munich]{Ludwig Maximilian University of Munich, Munich 80539, Germany}
\cortext[cor]{Corresponding author}

\begin{abstract}
\reva{This paper presents a computationally feasible method to compute rigorous bounds on the interval-generalisation of regression analysis to account for epistemic uncertainty in the output variables.} The new iterative method uses machine learning algorithms to fit an imprecise regression model to data that consist of intervals rather than point values. The method is based on a single-layer interval neural network which can be trained to produce an interval prediction. 
It seeks parameters for the optimal model that minimizes the mean squared error between the actual and predicted interval values of the dependent variable using a first-order gradient-based optimization and interval analysis computations to model the measurement imprecision of the data. \reva{An additional extension to a multi-layer neural network is also presented.}
We consider the explanatory variables to be precise point values, but the measured dependent values are characterized by interval bounds without any probabilistic information.
The proposed iterative method estimates the lower and upper bounds of the expectation region, which is an envelope of all possible precise regression lines obtained by ordinary regression analysis based on any configuration of real-valued points from the respective $y$-intervals and their $x$-values. 
\end{abstract}


\end{frontmatter}

\input{0-intro.tex}


\input{2-methodology}

\input{3-results}

\input{4-conclusions}

\input{5-appendix}

\bibliography{refs.bib}

\end{document}

%% file: 0-intro.tex
\newpage
\section{Introduction}

In real-life applications in science and engineering, data are often imperfect and contain uncertainty which arises, for example, from lack of knowledge, limited or inaccurate measurements, approximation or poor understanding of physical phenomena. This type of uncertainty is called \emph{epistemic uncertainty} or incertitude \cite{Oberkampf2001, epistemic_2007, epistemic_2009} and can be modelled, for example, by interval bounds which is the simplest method for projecting such uncertainty through mathematical expressions. This type of interval uncertainty can arise in various cases including non-detects and data censoring~\cite{Helsel2005}, periodic observations, plus-or-minus measurement uncertainties, interval-valued expert opinions \cite{SpeirsBridge2010}, privacy requirements, theoretical constraints, bounding studies, etc. \cite{SAND2007-0939, A_probabilistic_approach2011, Nguyen2012}. 

Interval data can be considered as an alternative to \say{symbolic data} modelled with uniform distributions over the interval ranges \cite{Uniform_approach2000, Billard_Diday2000, Billard_Diday2006}, which corresponds to Laplace’s principle of indifference with respect to the censored interval. All values within each interval are assumed to be equally likely. This assumption makes it relatively straightforward to calculate sample statistics for data sets containing such intervals. The alternative approach using only the ranges represents interval uncertainty purely in the form of bounds \cite{Manski2003, Xiang_Gang2006, SAND2007-0939}. This approach is consistent with the theory of imprecise probabilities \cite{Walley1991, Kuznetsov1991, Augustin2014}. It models each interval as a set of all possible values, and calculations result in a class of distribution functions, corresponding to the different possible values within the respective intervals.

Regression analysis problems have been extensively studied for many years, and special focus has been given to regression models with interval-valued data. In the setting of symbolic data analysis, Billard and Diday \cite{Billard_Diday2000} did pioneering work in which a linear regression model is fitted to the midpoint of the interval values. Over several decades various regression methods have been developed such as methods for linear regression models where the dependent variable has been either rounded or interval-censored \cite{Possibilistic_Approach_2011,SCHOLLMEYER2015}, methods for models that have interval-valued covariates and a precise dependent variable \cite{schollmeyer21a}, interval-based algorithms for the data-fitting problem where measurement errors are supposed to be bounded and known \cite{Shary2015,Kumkov2017}, some probabilistic \cite{Kreinovich_2016} and likelihood-based \cite{CATTANEO20121137} methods, non-parametric methods \cite{nonparametric_2013,Nasirzadeh2021}, and others \cite{Wang2012,BlancoFernndez2013}. \revb{In general, probabilistic models provide a more detailed uncertainty model, however interval-based techniques usually require fewer assumptions and provide theoretical guarantees on the reliability
and robustness of the results. The method proposed in this paper combines these two approaches.} 

Regression analysis using artificial neural networks is widely used due to the ability to learn the complex relationship between the explanatory variables and their response. Neural networks are capable of modeling complex systems, analyzing data and making predictions in a variety of research fields. Uncertainty quantification is essential in neural network predictions because of imperfect models and limited or imprecise data. Depending on the form of the data and how it was collected, uncertainty can be modelled by non-probabilistic or probabilistic methods.
\reva{Traditionally, it has been very common to make Gaussian assumptions about error structures.
Such methods include Bayesian neural networks \cite{BNN_1992,A_review_of_UQ2021}, variational inference \cite{Blei2017}, the Monte Carlo dropout technique \cite{Dropout_2014}.
Gaussian assumptions can be used in classification problems \cite{Zhou2018WeaklySI,9477300,NEURIPS2021} as well as regression problems.  For instance, linear regression residuals are classically assumed to be normally distributed.  However, this assumption of precise Gaussian distributions may not always be tenable for fully characterizing uncertainty when it includes both scatter and measurement imprecision.}

We propose a new iterative method for fitting an imprecise linear regression model to data that include epistemic uncertainties. The method is based on a single-layer interval neural network (SLINN) which can be trained to produce an interval prediction. An additional extension to a \revb{multi}-layer neural network is also presented. We consider the explanatory variables to be precise point values, and the dependent variables are presumed to have interval uncertainty. We assume that our imprecisely measured variables are surely enclosed within their respective censoring intervals, without necessarily knowing the distribution function characterizing the position of these values inside the intervals. \reva{Thus, the imprecision is modeled non-probabilistically even while the scatter of dependent values is modeled probabilistically by homoscedastic Gaussian distributions.} \reva{The main contributions of this paper are that it introduces a computationally inexpensive way to compute rigorous bounds on the interval-generalisation of regression analysis to account for epistemic uncertainty in the output (dependent) variables. The method escapes computational difficulties associated with intervals and generalizes the notion of a regression line. We compare this approach and review several related ideas in regressions that take account of epistemic uncertainty in the $y$-variable.}

\reva{Section \ref{Interval_Analysis} familiarizes with interval arithmetic and the dependency problem. Section \ref{Existing_methods} provides comparison with alternative strategies. Section \ref{Imprecise_Regression} describes the proposed  approach for imprecise linear regression with interval-valued $y$-data. Section \ref{Optimization_Algorithms} presents two first-order gradient-based optimization algorithms adapted to eliminate the dependency problem during interval computations. Sections \ref{forms_of_regression} and \ref{nn_app} consider nonlinear regressions and multi-layer interval neural networks. Section \ref{Numerical_Examples} applies the proposed methods to synthetic and real data sets, and \ref{Conclusion} offers conclusions.}

\section{Interval Analysis}\label{Interval_Analysis}

\subsection{Interval operations and functions}
An interval can be denoted
\begin{equation*}
    [x] = \interval{\underline{x}}{\overline{x}} = \{ x \in \mathbb{R} \mid \underline{x} \leq x \leq \overline{x} \} \in \mathbb{IR},
\end{equation*}
where $\underline{x}$ is the lower bound and $\overline{x}$ is the upper bound, and $\mathbb{IR}$ is the set of all intervals.
For intervals $\interval{\underline{x}}{\overline{x}}$ and $\interval{\underline{y}}{\overline{y}}$, the basic interval operations are defined by
\begin{equation}
    \begin{aligned}
        \interval{\underline{x}}{\overline{x}} + \interval{\underline{y}}{\overline{y}} &= \interval{\underline{x}+\underline{y}}{\overline{x}+\overline{y}}, \\
        \interval{\underline{x}}{\overline{x}} - \interval{\underline{y}}{\overline{y}} &= \interval{\underline{x}-\overline{y}}{\overline{x}-\underline{y}}, \\
        \interval{\underline{x}}{\overline{x}} \cdot \interval{\underline{y}}{\overline{y}} &= \interval{\min(\underline{x}\cdot\underline{y}, \underline{x}\cdot \overline{y},\overline{x}\cdot\underline{y},\overline{x}\cdot\overline{y}) }{\max(\underline{x}\cdot\underline{y}, \underline{x}\cdot \overline{y},\overline{x}\cdot\underline{y},\overline{x}\cdot\overline{y})}, \\
        \interval{\underline{x}}{\overline{x}} / \interval{\underline{y}}{\overline{y}} &= \interval{\underline{x}}{\overline{x}} \cdot 1/ \interval{\underline{y}}{\overline{y}} = \interval{\underline{x}}{\overline{x}} \cdot \left[ 1/\overline{y},1/\underline{y} \right] \quad \text{if $0\not\in \interval{\underline{y}}{\overline{y}}$}.
    \end{aligned}
\end{equation}
Additional operations for computing midpoint, width, and absolute value are defined as
\begin{equation}
    \begin{aligned}
        \m\interval{\underline{x}}{\overline{x}} & =  (\underline{x} + \overline{x})/2,\\
        \w\interval{\underline{x}}{\overline{x}} & = (\overline{x} - \underline{x}), \\
        |[x]| &= \max (|\underline{x}|,|\overline{x}|).
    \end{aligned}
\end{equation}

An $n$-interval vector $[\mathbf{x}] \in \mathbb{IR}^n$ is a vector with interval inputs
\begin{equation}
    [\mathbf{x}] = (\interval{\underline{x}_1}{\overline{x}_1}, \dots, \interval{\underline{x}_n}{\overline{x}_n}),
\end{equation}
and an $n \times m$ interval matrix $[\mathbf{X}] \in \mathbb{IR}^{n \times m}$ is a matrix with interval entries $[X_{ij}] = \interval{\underline{x}_{ij}}{\overline{x}_{ij}}$.\\

For a general real function $f: \mathbb{R}^n \rightarrow \mathbb{R}^m$ which maps an $n$-vector to an $m$-vector, the interval extension of $f$ is a function $\intsm{f}: \mathbb{IR}^n  \rightarrow \mathbb{IR}^m$ such that
\begin{equation}
    \intsm{f}(\intsm{\mathbf{x}}) \supseteq f(\intsm{\mathbf{x}}) = \{ f(\mathbf{y}) \mid \mathbf{y} \in \intsm{\mathbf{x}}\}.
\end{equation}
For monotonically increasing (or decreasing) functions including elementary functions such as $\log()$, $\exp()$, $\text{sqrt}()$ and others, the interval extension that produces sharp bounds can be computed using only endpoints
\begin{equation}
    f([x]) = \interval{\min\{f(\underline{x}),f(\overline{x})\} }{\max\{f(\underline{x}),f(\overline{x})\}}.
\end{equation}
An interval function $\intsm{f}$ is called inclusion monotonic 
if $[y] \subseteq [x]$ for $ \intsm{f}([y]) \subseteq \intsm{f}([x])$, where $[x],[y] \in \mathbb{IR}$.

\subsection{The dependency problem}
The dependency problem is a huge stumbling block in the efficient application of interval arithmetic which can lead to much wider bounds for the computed output than the exact range of the expression implies. 
The dependency problem happens when a mathematical expression includes multiple instances of an interval variable.
Naive replacement of floating point computations by intervals in an existing expression is not likely to lead to results that are best-possible. 
As an example, the interval-arithmetic computation of $[x] - [x]$ for $x \in [x] = \interval{1}{2}$ yields
\begin{equation*}
    [x] - [x] = \interval{-1}{1} 
\end{equation*}
which is true but arguably not best-possible 
if we hold that 
$x-x$ is zero no matter what possible values $x$ might have, in which case a better answer would be $[0,0] = 0$.

To overcome the dependency problem in interval computations a number of special-purpose algorithms have been developed to obtain best-possible results for a variety of particular scenarios where algebraically removing repetitions is impossible or problematic. 
This includes centered, slope and mean value forms \cite{Caprani1980, Moore2009},
affine arithmetic \cite{Rump1999,introductionAA},
Taylor models \cite{Taylor_Forms1983,Berz1998VerifiedIO},
Bernstein polynomials \cite{lorentz2013bernstein},
contractors, subpavings and other refinement strategies \cite{Caprani1975,Moore2009,Jaulin_2012}.

In this paper we will use the mean value form which is a second-order inclusion monotone approximation for an interval-valued extension of the real function $f(x)$
\begin{equation}\label{mvf}
    f([x]) \subseteq f_m([x]):= f(\m [x] ) + \mathbb{J}_f([x]) \left([x] - \m [x]\right),
\end{equation}
where $\mathbb{J}_f([x])$ is an interval inclusion of the Jacobian matrix of the $f$ function. In the case of an $m \times 1$ interval vector $[\mathbf{x}]$ the second term in (\ref{mvf}) reads $\sum_{i=1}^m D_i f ([x_i] - \m [x_i])$, where $D_i f$ is an interval extension of the partial derivatives $\partial f/ \partial x_i $.

\section{\revb{Related Work}}\label{Existing_methods}

\textit{Classical regression methods.} \revb{Billard Diday \cite{Billard_Diday2000} proposed the center method (CM) which deals with interval-valued data in linear regression. This approach only takes into account the center points of the intervals, while leaving out other important information such as the widths of the intervals. Subsequent improvements to this method had been proposed in \cite{Carvalho_2004,Billard_Diday2006,LIMANETO_2010,Xu_2010} including separate regression models for the center and the range of the intervals, constrained linear regression models, and models based on the symbolic sample covariance. The center and range method (CRM) proposed by De Carvalho and Lima Neto \cite{Carvalho_2004} estimates the parameters for a linear model using both the center points and widths of the ranges of the interval-valued data as two independent models. The constrained center and range method (CCRM) \cite{LIMANETO2010_CCMR} mathematically ensures that the lower bound is less or equal to the upper bound.}

\textit{The sharp collection region.} The methodology of imprecise probabilities offers powerful methods for reliable handling of coarse data \cite{Walley1991, horowitz2001imprecise}, which includes all situations where data are not observed in perfect resolution. The proposed iterative approach can be compared to an alternative strategy for handling interval data in the dependent variable of regression models. The sharp collection region (SCR) is one of three types of so-called \say{identification regions} that are discussed by Schollmeyer and Augustin in~\cite{SCHOLLMEYER2015}, which are sets of all parameters compatible with the interval data and model. This collection region has been studied, e.g., in \cite{beresteanu,beresteanub,Possibilistic_Approach_2011} (cf., also \cite{ponomareva}).

Let $\Theta$ be a parameter space and $P = \{ \mathbb{P}_\theta \mid \theta \in \Theta \}$ 
be a statistical model associated with the unobserved random variable $y$ and the three observable random variables $x,\underline{y},\overline{y}$, assuming $y \in [\underline{y}, \overline{y}]$. With $\mathbb{P}$ we denote the unknown true model and with $\mathbb{E}$ the corresponding expectations. The sharp collection region is defined as
\begin{equation}
    SCR:= \bigcup\limits_{z \in \mathbb{E}([\underline{y},\overline{y}] \mid x)} \argmin_{\theta \in \Theta} L \left( F^{x,y}_\theta, F^{x,z} \right),
\end{equation}
where $F^{x,y}_\theta$ is a joint distribution of the random variables under the $\mathbb{P}_\theta$ model,  $F^{x,z}$ is the joint distribution under a possible true model, and $L$ is a loss function which is zero if and only if the arguments are equal. The range of the expectation $\mathbb{E}([\underline{y},\overline{y}] \mid x)$ denotes the set of all values $z$ fulfilling the condition $\mathbb{E}(\underline{y}\mid x) \leq \mathbb{E}(z \mid x) \leq \mathbb{E}(\overline{y} \mid x)$. The sharp collection region is the collection of all models that describe a possible random variable $z$ consistent with the boundaries $\underline{y}, \overline{y}$ in the best way in terms of the loss function $L$. For a linear model and the classical quadratic loss function we have
\begin{equation}
    SCR(\underline{y},\overline{y}) = \{\argmin \mathbb{E}((\beta_0 + \beta_1 x - y)^2) \mid y \in [\underline{y},\overline{y}]\}.
\end{equation}
It actually turns out that for the special case of linear regression under quadratic loss, the collection region is identical to the SLINN method. However, the SLINN approach is far more flexible and can be more widely adapted, e.g., to non-linear regression. A comparison of the proposed method, the sharp collection region and the center and range methods for a linear regression  is presented in Section \ref{Numerical_Examples}.

\textit{Monte Carlo simulations.} We also compare the proposed single-layer interval neural network to a brute-force strategy based on Monte Carlo simulations which simply wiggles the points within the respective $Y$-intervals and searches for extreme cases. The most serious limitation of Monte Carlo simulations is that it represents all uncertainties as random variables. For example, in a large-scale problem which has many uncertain variables (like imprecise regression does), or which is computationally expensive to evaluate, the cost of Monte Carlo computations can become prohibitive. However, there are a number of special algorithms such as low-discrepancy sequences and Latin-hypercube sampling that can speed up the convergence of Monte Carlo.

\textit{Neural networks.} The first ideas to incorporate interval-valued data in neural networks were described by Ishibuchi et al. \cite{Ishibuchi_1991} and Hernandez et al. \cite{Interval_backpropagation_1993} in training feed-forward neural networks and extending the backpropagation learning algorithm. \revb{Additional improvements to this strategy in the context of fuzzy regression analysis and nonlinear interval models were reported in \cite{Ishibuchi_1993,Huang1998}. Traditional interval neural networks such as \cite{Ishibuchi_1993} and other versions mentioned above are based on interval arithmetic, but the authors apply only three interval operations (scalar multiplication and addition and multiplication of intervals) and avoid the subtraction operation. Thus their loss function which we denote $L_{\text{ISH}}$ \cite{Ishibuchi_1993} is not an interval and does not require general interval computations as it is defined as a sum of the differences between the upper and lower bounds separately. We believe that such a definition is valid under specific conditions, however only the use of pure interval arithmetic can guarantee that the obtained results are rigorous.}

\revb{Other recent developments of the artificial neural networks for interval-valued data prediction such as \cite{INN_regression_Chetwynd2006, Roque2007,Yang_INN2018,Yang_INN2019} are based on the CM and the CRM methods for linear regression. These neural networks exploit the loss function which is a weighted Euclidean distance between the predicted intervals $[h(x_i)]$ and observed intervals $[y_i]$ such that
\begin{equation}
   \gamma (\m[h(x_i)]  - \m[y_i])^2 + (1-\gamma)(\w[h(x_i)]/2  - \w[y_i]/2)^2,
\end{equation}
where $\gamma \in [0,1]$ is the weighting parameter which measures the relative importance of center points and radii. The use of this loss function is somewhat similar to Ishibuchi's method in terms of avoiding an intervalized loss function and direct subtraction operations. In addition there is the bother of specifying the constant $\gamma$ which may affect the results in an unclear way.}

In~\cite{Sadeghi2019} Sadeghi et al.\  presented an alternative way to efficiently train interval neural networks for imprecise data using the framework of interval predictor models~(IPM). \revb{The IPM \cite{ipm_2009, Linf_layer2009} maps some observable variables (system inputs) into an interval that is used to predict an inaccessible variable. The regression model is tuned according to the minimax criterion of best fit, which amounts to selecting the parameters that minimize the maximum deviation from the observed $y_i$ by $h(x_i)$, i.e. \(\max\limits_{i}\) $ |y_i - h(x_i)|$. Unlike  standard regression techniques, these models try to bound an envelope of the data using a conceptualization of fit different from the traditional least squares criterion, which produce interval estimates for the regression even from point-valued input data.}

\revb{Summarizing the above, for the case of interval imprecision in the output variables with pairs $x_i, \intsm{Y_i} = \interval{\underline{y}_i}{\overline{y}_i}$ we have
\begin{equation*}
    \begin{aligned}
        L_{\text{INN}}&= \dfrac{1}{2}\norm{[h] - [Y]}^2, \\
        L_{\text{ISH}}&= \dfrac{1}{2}\norm{\underline{h} - \underline{Y}}^2 + \dfrac{1}{2}\norm{\overline{h} - \overline{Y}}^2, \\
        L_{\text{IPM}}&= \max_{i} \max(|h(x_i) - \underline{y}_i|,|h(x_i) - \overline{y}_i|),
    \end{aligned}
\end{equation*}
where $L_{\text{ISH}}$ is the loss function for the traditional approach \cite{Ishibuchi_1993}, $L_{\text{IPM}}$ is the loss function for Sadeghi's interval predictor model \cite{Sadeghi2019}, and $L_{\text{INN}}$ is our loss function. Several comparative examples are presented in Section \ref{Numerical_Examples}.} \reva{The loss functions proposed in this paper apply to regression problems. No attempt has been made to generalise them for other problems like classification or cross entropy.}

%% file: 2-methodology.tex
\section{Imprecise Regression}\label{Imprecise_Regression}

\subsection{Statistical model and caveats}

We consider a general linear model with one dependent variable where  $\by $ is a $n \times 1$ vector of observations on the dependent variable, $\bX$ is a $n \times m$ augmented matrix with $m$ independent variables for $n$ observations including a constant term which is represented by column of ones, $\epsilon$ is an $n \times 1$ vector of random Gaussian errors, and $\beta$ is a $m \times 1$ vector of parameters to be estimated from the regression model
\begin{equation}\label{our_model}
    \begin{bmatrix}
        y_1 \\
        y_2 \\
        \vdots \\
        y_n 
    \end{bmatrix} 
    =
    \begin{bmatrix}
        1 & x_{11} & \cdots & x_{1m}\\
        1 & x_{21} & \cdots & x_{2m}\\
        \vdots & \vdots & \ddots & \vdots \\
        1 & x_{n1} & \cdots & x_{nm}
    \end{bmatrix}
    \begin{bmatrix}
        \beta_0 \\
        \beta_1 \\
        \vdots \\
        \beta_m 
    \end{bmatrix}
    +
    \begin{bmatrix}
        \epsilon_1 \\
        \epsilon_2 \\
        \vdots \\
        \epsilon_n 
    \end{bmatrix} 
\end{equation}
or, in matrix form,
\begin{equation}
    \by = \bX \beta + \epsilon.
\end{equation}
In order to obtain a vector of regression coefficients $\beta$ we use ordinary least squares estimation which minimizes the sum of squared residuals
\begin{equation}
    L = \sum_{i=1}^n \left(y_i -  \beta_0 + \sum_{j=1}^m x_{ij} \beta_j \right)^2 = \norm{ \by - \bX \beta}^2,
\end{equation}
and yields a closed-form solution
\begin{equation}\label{closed_form_solution}
    \beta = (\bX^T\bX)^{-1}\bX^T \by.
\end{equation}

\begin{figure}[ht!]
	\centering
	\begin{subfigure}{0.49\textwidth}
		\includegraphics[width=\linewidth]{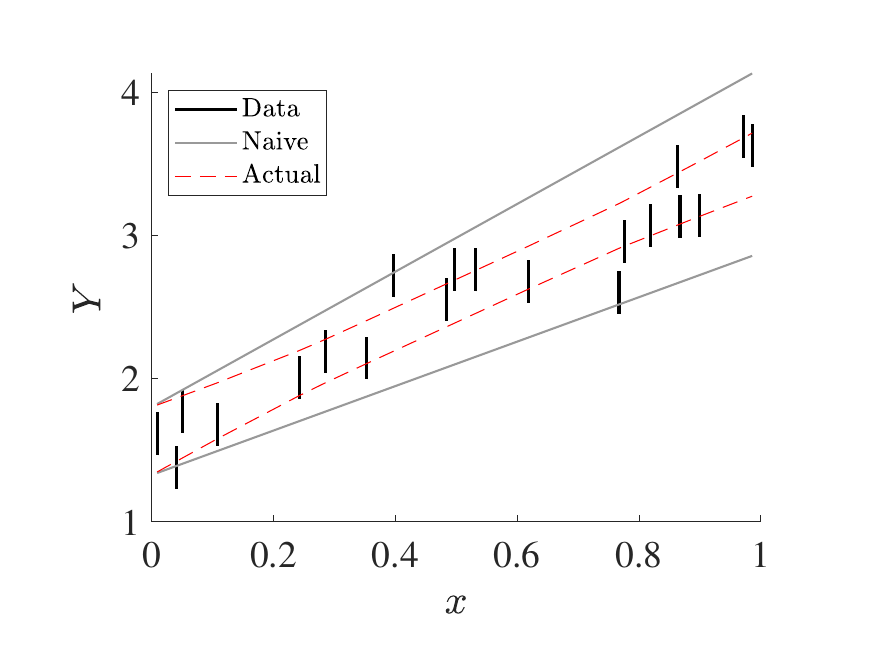}
		\caption{}
	\end{subfigure}
	\begin{subfigure}{0.49\textwidth}
		\includegraphics[width=\linewidth]{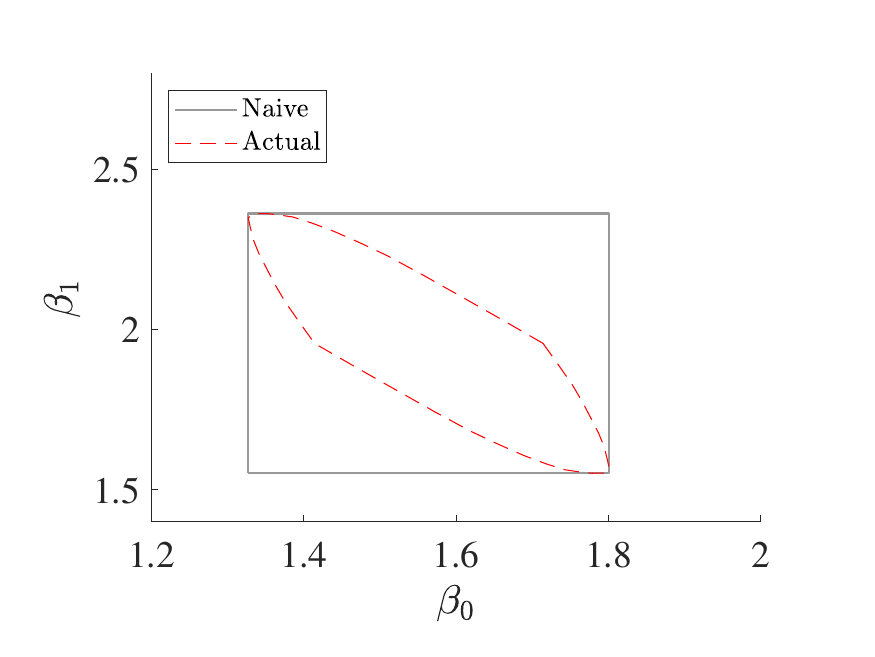}
		\caption{}
	\end{subfigure}
	\caption{An example of imprecise regression: (a) computed expectation region (grey lines) using naive approach and \revb{best-possible bounds} (red dashed lines) on all possible lines from regressions based on any possible configuration of points; (b) rectangle for all possibles values (in grey) \revb{and the identification region (red dashed lines)}.}
	\label{fig:naive_approach}
\end{figure}

When the dependent values in $\by$ have uncertainty, they can each be replaced by interval bounds $\intsm{Y_i} = \interval{\underline{y}_i}{\overline{y}_i}$ that guarantee the true value lies within them. 
In this case equation (\ref{closed_form_solution}) becomes
\begin{equation}\label{scottwhines}
    \interval{\underline{\beta}}{\overline{\beta}} = (\bX^T\bX)^{-1}\bX^T \intsm{Y},
\end{equation}
which gives the estimated interval coefficients $\interval{\underline{\beta}}{\overline{\beta}}$. In the case of simple linear regression ($m=1$), a naive solution would lead to these overly wide results  
\begin{equation}\label{naive_regression}
    [\hat{Y}_i] = \interval{\underline{\beta}_0}{\overline{\beta}_0} + \interval{\underline{\beta}_1}{\overline{\beta}_1} x_{i1},
\end{equation}
because the interval $Y$-values each appear two times in the expression for the $\beta$-values. This naive approach uses a single step to solve closed-form equation (\ref{scottwhines}) and forms the rectangle for all possibles values with bounds  $\interval{\underline{\beta}_0}{\overline{\beta}_0}, \interval{\underline{\beta}_1}{\overline{\beta}_1}$. An illustrative example is presented in Fig. \ref{fig:naive_approach} where solid grey lines correspond to the \revb{naive} approach and red lines show the \revb{best-possible bounds considering all possible configurations of the $y$-data within their respective interval as an} expectation region \reva{which is a generalization of a regression line}. \revb{We call these best-possible bounds the `actual' bounds.}
\revb{The identification region depicted within the red dashed lines in Fig.~\ref{fig:naive_approach}b was computed using an algorithm based on the insight that this region is the image of a multidimensional cuboid under a linear map (i.e., a zonotope, cf., \cite{Possibilistic_Approach_2011}) for the sharp collection region \cite{SCHOLLMEYER2015}.}

It is important to note that there is a functional dependence between the slope and intercept which fills only a proper subset of the rectangle. Not all combinations are allowed.
An optimal generalization of regression analysis to interval data would instead identify the \emph{envelope} of all possible lines from regressions based on any possible configuration of points within the respective intervals $Y_i$.  All such lines use only permissible combinations of $\beta_0$ and $\beta_1$.
As Fig.~\ref{fig:naive_approach} suggests, the separate intervals for the slope and intercept are correct, but ignoring their dependence yields unnecessarily wide answers.

\subsection{Iterative approach}

The proposed iterative method is based on machine learning algorithms which fit an imprecise linear regression model to observations that include interval values. 
The regression model assumes that $\mathbb{E}([Y] | \bX)$ is linear in the unknown interval parameters to be estimated from the data. 
It generalises a single-layer neural network consisting of an input layer $\bX$ and output layer $h$ which replaces $\hat{Y}$ as the prediction. Generally, any layer of a neural network can be considered as an affine transformation with subsequent application of an activation function $F$ forming
\begin{equation*}
    \textit{Input}: \bX \quad \quad \textit{Output}: h = F(\bX \bW + \bb),
\end{equation*}
where $\bW$ is a weight matrix and $\bb$ is a column-vector containing bias terms.

The proposed iterative method is a single-layer interval neural network, a supervised learning algorithm in a regression setting with interval-valued dependant variables. In the context of linear regression an activation function $F$ is not needed, and the SLINN iteratively optimises $\bW$ and $\bb$ by minimizing the interval mean squared error loss function 
\begin{equation}\label{loss_function}
    L = \dfrac{1}{2N} \norm{h - [Y]}^2 = \dfrac{1}{2N} \norm{\bX \bW + \bb - [Y]}^2,
\end{equation}
which is assumed to be defined and differentiable with respect to both $\bb$ and $\bW$. 

Training the interval extension of a real-valued neural network includes the following major steps:
\begin{enumerate}
    \item performing forward propagation to calculate an interval prediction $h$,
    \item computing the loss function at each training iteration which quantifies the distance between the actual $Y$ and predicted values $h$,
    \item computing the gradient of the loss function with respect to interval parameters,
    \item minimizing the loss function with gradient descent or another optimization technique.
\end{enumerate}
To perform interval arithmetic computations during the training process we use the specialized Matlab toolbox INTLAB~\cite{Rump1999} which provides powerful and rigorous tools for matrix operations. In this paper we consider two first-order gradient-based optimization algorithms and adapt them to interval computations eliminating the dependency problem.

\section{Optimization Algorithms}\label{Optimization_Algorithms}

\subsection{Gradient descent}\label{Gradient_descent}
We shall first consider the most straightforward and popular first-order iterative optimization algorithm known as gradient descent \cite{Courant1943, gd}. 
The algorithm updates the parameters adaptively at each iteration $k$ in the direction of the negative gradient of the loss function 
\begin{align}
    \bW^{(k+1)} &= \bW^{(k)} - \alpha \nabla_{\bW^{(k)}} L, \label{GD_updateW} \\
    \bb^{(k+1)} &= \bb^{(k)} - \alpha \nabla_{\bb^{(k)}} L, \label{GD_updateb}
\end{align}
where $\alpha \in \mathbb{R}^{+}$ is the learning rate which controls the size of the steps we take to reach minima.

Interval operations provide results that are rigorous in the sense that they bound all possible answers. The results are best-possible when they cannot be any tighter without excluding some possible answers. However, a direct naive substitution of the interval vector $Y$ to the loss function (\ref{loss_function}) will cause the dependency problem as discussed in Section~\ref{Interval_Analysis}. The updating rules (\ref{GD_updateW}) and~(\ref{GD_updateb}) add extra instances of intervals on each iteration to $\bW^{(k+1)}(\bW^{(k)},\bb^{(k)},Y)$, $\bb^{(k+1)}(\bW^{(k)},\bb^{(k)},Y)$ which results in inflated bounds. 
Here and below we omit the use of square brackets for interval values $\bW,\bb,Y$ and output $h$ for simplicity.
Consider, for example, the sequence of iterations for updating the weights  
\begin{equation}
    \begin{matrix}
        \bW^{(1)} &=& \bW^{(0)} - \alpha \bX^T (\bX \bW^{(0)} + \bb^{(0)} - Y), \\
        \bW^{(2)} &=& \bW^{(1)} - \alpha \bX^T (\bX \bW^{(1)} + \bb^{(1)} - Y), \\
        \vdots && \vdots\\
        \bW^{(k+1)} &=& \bW^{(k)} - \alpha \bX^T (\bX \bW^{(k)} + \bb^{(k)} - Y).
    \end{matrix}
\end{equation}
At any $(k+1)$ iteration $\bW^{(k+1)}$ is the linear function
\begin{equation}\label{W_linearfun}
    \bW^{(k+1)} = f(\bW^{(k)}(\dots (\bW^{(0)},\bb^{(0)},Y)), \; \bb^{(k)}(\dots (\bW^{(0)},\bb^{(0)},Y)), \; Y)
\end{equation}
with respect to initial $\bW^{(0)},\bb^{(0)}$ and $Y$. In order to overcome the dependency problem in the iterative approach, we use the mean value form to approximate the interval-valued function (\ref{W_linearfun}).  
The key idea is to apply form~(\ref{mvf}) to equation~(\ref{GD_updateW}) so that for any
interval variable $u =  \interval{\underline{u}}{\overline{u}}$ we have
\begin{equation}\label{mvf_W_grad_descent}
    \bW^{(k+1)}(u) = \bW^{(k+1)}(\m u) + \mathbb{J}_{\bW}(u) (u - \m u).
\end{equation}
In more general cases when $Y$ and the initial guesses $\bW^{(0)}, \bb^{(0)}$ are intervals equation (\ref{mvf_W_grad_descent}) can be written in expanded form
\begin{multline}\label{mvf_expanded_grad}
        \bW^{(k+1)}(\bW^{(0)},\bb^{(0)},{Y}) = \bW^{(k+1)}(\m{\bW^{(0)}},\m{\bb^{(0)}},\m{Y} ) + \\
    + \begin{bmatrix}
        \dfrac{\dr \bW^{(k+1)}}{\dr \bW^{(0)}} & \dfrac{\dr \bW^{(k+1)}}{\dr \bb^{(0)}}  \dfrac{\dr \bW^{(k+1)}}{\dr Y} 
    \end{bmatrix}
    \begin{bmatrix}[1.25]
    {\bW^{(0)}} - \m {\bW^{(0)}} \\
    {\bb^{(0)}} - \m {\bb^{(0)}} \\
    {Y} - \m {Y}
    \end{bmatrix}.
\end{multline}
The first term in (\ref{mvf_expanded_grad}) is nothing more than the standard updating rule for $\bW$ with precise midpoint values. 
Because the equation (\ref{W_linearfun}) is a linear function, the mean value form yields best-possible results even though it has multiple repeated interval variables, so
the Jacobian matrix has only precise values without any uncertainty.

All three derivatives at iteration $(k+1)$  can be computed using the chain rule
\begin{align}
    \dfrac{\dr \bW^{(k+1)}}{\mathrm{d} Y}  &= \dfrac{\partial \bW^{(k+1)}}{\partial \bW^{(k)}} \dfrac{\partial \bW^{(k)}}{\partial Y} + \dfrac{\partial \bW^{(k+1)}}{\partial \bb^{(k)}} \dfrac{\partial \bb^{(k)}}{\partial Y}  + \dfrac{\partial \bW^{(k+1)}}{\partial Y}, \label{dWdY} \\
    \dfrac{\dr \bW^{(k+1)}}{\dr \bW^{(0)}}  &= \dfrac{\partial \bW^{(k+1)}}{\partial \bW^{(k)}} \dfrac{\partial \bW^{(k)}}{\partial \bW^{(0)}} + \dfrac{\partial \bW^{(k+1)}}{\partial \bb^{(k)}} \dfrac{\partial \bb^{(k)}}{\partial \bW^{(0)}}, \\
    \dfrac{\dr \bW^{(k+1)}}{\dr \bb^{(0)}}  &= \dfrac{\partial \bW^{(k+1)}}{\partial \bW^{(k)}} \dfrac{\partial \bW^{(k)}}{\partial \bb^{(0)}} + \dfrac{\partial \bW^{(k+1)}}{\partial \bb^{(k)}} \dfrac{\partial \bb^{(k)}}{\partial \bb^{(0)}}.
\end{align}
The same argument applies to the biases $\bb^{(k+1)}$ and their derivatives from the Jacobian. 
Taking formula (\ref{dWdY}) as example, notice that the first multiplier in the first term is a constant coefficient for any $k$. We call this structure $\partial a^{(k+1)} / \partial b^{(k)} = \partial a^{(1)} / \partial b^{(0)}$ a \emph{neighboring derivative} which needs to be computed only once at the beginning of an iterated sequence. The second multiplier in the first term comes from the previous iteration. 
The same logic about the neighboring derivative applies to the second term in (\ref{dWdY}), where the second multiplier comes from the $k$th result of the $\bb$ iteration.
These interlinked iterations can be easily computed because the linearity of the first-order structure of the optimisation algorithm reveals each next derivative to be a combination of the derivative from the previous iteration multiplied by the neighboring derivative which is constant for any $k$, across the linked iterations.

Although 
$\bW^{(k)}$ and $\bb^{(k)}$ can be optimally evaluated for any $k$, 
naively computing the interval output layer $h^{(k+1)} = \bX \bW^{(k)} + \bb^{(k)}$  also yields inflated uncertainty because of the unseen dependence between $\bW^{(k)}$ and $\bb^{(k)}$ caused by the repeated  $Y$-values.  To overcome this, the mean value form can be used for $h$ as well.  The relevant derivatives are
\begin{align}
    \dfrac{\dr h^{(k+1)}}{\dr Y } &= \bX \dfrac{\dr \bW^{(k)}}{\dr Y} + \dfrac{\dr \bb^{(k)}}{\dr Y}, \label{h_deriv}
\end{align}
and the resulting mean value form expression is
\begin{multline}\label{mvf_layer}
        h^{(k+1)}(\bW^{(0)},\bb^{(0)},{Y}) = h^{(k+1)}(\m{\bW^{(0)}},\m{\bb^{(0)}},\m{Y} ) + \\
    + 
    \begin{bmatrix}
        \dfrac{\dr h^{(k+1)}}{\dr \bW^{(0)}} & \dfrac{\dr h^{(k+1)}}{\dr \bb^{(0)}} & \dfrac{\dr h^{(k+1)}}{\dr Y }
    \end{bmatrix}
    \begin{bmatrix}[1.25]
    {\bW^{(0)}}- \m {\bW^{(0)}} \\
    {\bb^{(0)}} - \m {\bb^{(0)}} \\
    {Y} - \m {Y}
    \end{bmatrix}.
\end{multline}
The first term in (\ref{mvf_layer}) is the precise central line fitted to midpoints taken from the interval inputs.  We denote it as $\m h^{(k+1)}$ which will be useful in Section~\ref{nn_app}. Equation~(\ref{mvf_layer}) can be thus written more compactly as 
\begin{equation}\label{mvf_line_shortform}
     h^{(k+1)}(\bW^{(0)},\bb^{(0)},{Y}) = \m h^{(k+1)} + \mathbb{J}_h     
    \begin{bmatrix}[1.25]
    \bW^{(0)}- \m \bW^{(0)} \\
    \bb^{(0)} - \m \bb^{(0)} \\
    Y - \m Y
    \end{bmatrix}.
\end{equation}

\subsection{Gradient descent with momentum}

Gradient descent is known to be slow compared to other optimization algorithms and can converge to a local minimum or saddle point when the loss function is highly non-convex. 
Choosing too small a learning rate leads to terribly slow convergence, while large $\alpha$ can lead to divergence or to oscillating effects across the slope of the ravine.
Also the learning process can be slow when an update is performed through all observations.

As an alternative, we adapt another optimization algorithm which is very similar to gradient descent but helps to accelerate it in the relevant direction. Gradient descent  \emph{with momentum} keeps in memory previous updates and has the following update rules:
\begin{align}
    \bW^{(k+1)} &= \bW^{(k)} - \alpha \dw^{(k+1)}, \label{mom_dW}\\
    \bb^{(k+1)} &= \bb^{(k)} - \alpha \db^{(k+1)}, \label{mom_db}\\
        \dw^{(k+1)} &= (1-\beta) \nabla_{\bW^{(k)}}L + \beta \dw^{(k)}, \\
    \db^{(k+1)} &= (1-\beta) \nabla_{\bb^{(k)}}L + \beta \db^{(k)},
\end{align}
where $\beta$ is a hyperparameter called momentum that takes a value from 0 to 1.
This momentum method uses an exponentially weighted average for $\nabla_{\bW^{(k)}}L$ and $\nabla_{\bb^{(k)}}L$ values and smooths down gradient measures. In order to adapt the momentum method to interval computations and reduce the interval dependency we rewrite equations (\ref{mom_dW}-\ref{mom_db}) as
\begin{equation}\label{Wb_expanded}
\begin{aligned}
    \bW^{(k+1)} &= \bW^{(k)} - \alpha (1-\beta) \left[ \bS^{(k)} + \beta \bS^{(k-1)} + \beta^2 \bS^{(k-2)} + \dots + \beta^{k} \bS^{(0)}\right], \\
    \bb^{(k+1)} &= \bb^{(k)} - \alpha (1-\beta) \left[ \bD^{(k)} + \beta \bD^{(k-1)} + \beta^2 \bD^{(k-2)} + \dots + \beta^{k} \bD^{(0)}\right],
\end{aligned}
\end{equation}
and define the intermediate functions $\bS$, $\bD$
\begin{equation}
    \begin{aligned}
    \bS^{(k)}(\bW^{(k)}(Y),\bb^{(k)}(Y),Y) &= \bX^T (\bX \bW^{(k)} + \bb^{(k)} - Y), \\
    \bD^{(k)}(\bW^{(k)}(Y),\bb^{(k)}(Y),Y) &= (\bX \bW^{(k)} + \bb^{(k)} - Y).
    \end{aligned}
\end{equation}

As in Section~\ref{Gradient_descent} we use the mean value form to approximate the interval functions $\bW$ and $\bb$, which requires computing derivatives of $\bW$ and $\bb$ in (\ref{Wb_expanded}) with respect to interval values. We can assume, for simplicity, that our initial guesses for $\bW^{(0)}, \bb^{(0)}$ are certain values, so formula (\ref{mvf_expanded_grad}) becomes a bit easier for this case:
\begin{equation}\label{mvf_momentum}
\begin{aligned}
        \bW^{(k+1)}(Y) = \bW^{(k+1)}(\m Y) + \dfrac{\dr \bW^{(k+1)}}{\dr Y} \left({Y} - \m {Y}\right).
\end{aligned}
\end{equation}
Taking the full derivative of the multivariable function $\bW^{(k+1)}$ with respect to the interval vector $Y$ for $k=2,3,4$ iterations yields
\begin{align}
\begin{split}
\dfrac{\dr \bW^{(2)}}{\dr Y}  &= \dfrac{\partial \bW^{(2)}}{\partial Y} + \dfrac{\partial \bW^{(2)}}{\partial \bW^{(1)}} \dfrac{\partial \bW^{(1)}}{\partial Y} + \dfrac{\partial \bW^{(2)}}{\partial \bb^{(1)}} \dfrac{\partial \bb^{(1)}}{\partial Y},
\end{split}
\\[2ex]
\begin{split}\label{it_3}
        \dfrac{\dr \bW^{(3)}}{\dr Y}  &= \dfrac{\partial \bW^{(3)}}{\partial Y} + \dfrac{\partial \bW^{(3)}}{\partial \bW^{(2)}} \dfrac{\partial \bW^{(2)}}{\partial Y} + \dfrac{\partial \bW^{(3)}}{\partial \bb^{(2)}} \dfrac{\partial \bb^{(2)}}{\partial Y} + \\ 
        &+ \dfrac{\partial \bW^{(3)}}{\partial \bW^{(1)}} \dfrac{\partial \bW^{(1)}}{\partial Y}  + \dfrac{\partial \bW^{(3)}}{\partial \bb^{(1)}} \dfrac{\partial \bb^{(1)}}{\partial Y},
\end{split}
\\[2ex]
\begin{split}\label{it_4}
        \dfrac{\dr \bW^{(4)}}{\dr Y}  &= \dfrac{\partial \bW^{(4)}}{\partial Y} + \dfrac{\partial \bW^{(4)}}{\partial \bW^{(3)}} \dfrac{\partial \bW^{(3)}}{\partial Y} + \dfrac{\partial \bW^{(4)}}{\partial \bb^{(3)}} \dfrac{\partial \bb^{(3)}}{\partial Y} + \\
        &+ \dfrac{\partial \bW^{4)}}{\partial \bW^{(2)}} \dfrac{\partial \bW^{(2)}}{\partial Y} + \dfrac{\partial \bW^{(4)}}{\partial \bb^{(2)}} \dfrac{\partial \bb^{(2)}}{\partial Y} + \\
        &+ \dfrac{\partial \bW^{(4)}}{\partial \bW^{(1)}} \dfrac{\partial \bW^{(1)}}{\partial Y} + \dfrac{\partial \bW^{(4)}}{\partial \bb^{(1)}} \dfrac{\partial \bb^{(1)}}{\partial Y},
\end{split}
\end{align}
so for iteration $(k+1)$ we have
\begin{equation}\label{dW_dY_momentum}
    \begin{split}
        \dfrac{\dr \bW^{(k+1)}}{\dr Y}  &= \dfrac{\partial \bW^{(k+1)}}{\partial Y} + \dfrac{\partial \bW^{(k+1)}}{\partial \bW^{(k)}} \dfrac{\partial \bW^{(k)}}{\partial Y} + \dfrac{\partial \bW^{(k+1)}}{\partial \bb^{(k)}} \dfrac{\partial \bb^{(k)}}{\partial Y} + \\
        &+ \dfrac{\partial \bW^{(k+1)}}{\partial \bW^{(k-1)}} \dfrac{\partial \bW^{(k-1)}}{\partial Y} + \dfrac{\partial \bW^{(k+1)}}{\partial \bb^{(k-1)}} \dfrac{\partial \bb^{(k-1)}}{\partial Y} + \hdots +\\
        &+ \dfrac{\partial \bW^{(k+1)}}{\partial \bW^{(1)}} \dfrac{\partial \bW^{(1)}}{\partial Y} + \dfrac{\partial \bW^{(k+1)}}{\partial \bb^{(1)}} \dfrac{\partial \bb^{(1)}}{\partial Y}.
\end{split}
\end{equation}
The number of terms in (\ref{dW_dY_momentum}) increases with each iteration because
the updating rules for momentum include the functions $\bS$ and $\bD$ from previous iterations unlike the standard gradient descent. 
However, expression (\ref{dW_dY_momentum}) depends on the previously computed $\partial \bW / \partial Y$ and  $\partial \bb / \partial Y$ and has same neighboring derivatives as in formula (\ref{dWdY}) which should be computed once. In addition, we use the relationships between the derivatives
\begin{equation}\label{relations}
     \dfrac{\partial \bW^{(k+1)}}{\partial \bW^{(k-2)}} = \beta \dfrac{\partial \bW^{(k)}}{\partial \bW^{(k-2)}},  \quad      \dfrac{\partial \bW^{(k+1)}}{\partial \bb^{(k-2)}} = \beta \dfrac{\partial \bW^{(k)}}{\partial \bb^{(k-2)}}, \quad \text{for} \quad k \geq 3.
\end{equation}
Thus, the two last terms in (\ref{it_4}) are nothing more than the two last terms from (\ref{it_3}) multiplied by $\beta$. Assuming (\ref{relations}) and the sequential dependency on previous iterations, it is possible to rewrite (\ref{dW_dY_momentum}) in a more compact form
\begin{equation}
    \dfrac{\dr \bW^{(k+1)}}{\dr Y}  = \dfrac{\partial \bW^{(k+1)}}{\partial Y} + \dfrac{\partial \bW^{(k+1)}}{\partial \bW^{(k)}} \dfrac{\partial \bW^{(k)}}{\partial Y} + \dfrac{\partial \bW^{(k+1)}}{\partial \bb^{(k)}} \dfrac{\partial \bb^{(k)}}{\partial Y}  + \mathbb{M}^{(k)}
\end{equation}
where $\mathbb{M}^{(1)} = 0$, and for $k \geq 2$
\begin{equation}
    \mathbb{M}^{(k)} = \dfrac{\partial \bW^{(k+1)}}{\partial \bW^{(k-1)}} \dfrac{\partial \bW^{(k-1)}}{\partial Y} + \dfrac{\partial \bW^{(k+1)}}{\partial \bb^{(k-1)}} \dfrac{\partial \bb^{(k-1)}}{\partial Y} + \beta \; \mathbb{M}^{(k-1)}.
\end{equation}

The particular derivatives needed for the linear and nonlinear regression settings are shown in~\ref{app}.

\section{Other Forms of Nonlinear Regression  Analysis}\label{forms_of_regression}

The model $\mathbb{E}( [Y] | X)$ that we use here is linear with respect to unknown interval parameters, but not necessarily in the input variables $X$. Therefore, it is possible to augment the model by replacing the input variables with nonlinear basis functions of the input variables for the one-dimensional case such as $\sum_{j=0}^m w_j \phi_j(x)$,
where the $w_j$ are the regression coefficients. One of the possible options for a basis is to set $\phi_j(x) = x^j$, where $j=0,1,\dots,m$. Such functional relationship is called polynomial regression, which is considered to be a special case of multiple linear regression. For this case the general design matrix in~(\ref{our_model}) can be rewritten as
\begin{equation}
    \bX = \begin{bmatrix}
        1 & \phi_1(x_1) & \phi_2(x_1) & \dots & \phi_m(x_1)\\
        1 & \phi_1(x_2) & \phi_2(x_2) & \dots & \phi_m(x_2)\\
        \vdots & \vdots & \vdots & \ddots & \vdots \\
        1 & \phi_1(x_n) & \phi_2(x_n) & \dots & \phi_m(x_n)\\
    \end{bmatrix}.
\end{equation}
Because the matrix $\bX$ is constant and doesn't contain any uncertain variables, this makes it straightforward to compute the imprecise regression model with polynomial basis functions using the SLINN as $[h] = \bX[\bW]$, where $[\bW_j] = \interval{\underline{w}_j}{\overline{w}_j}$.
This expression does not have bias (intercept) term because we are using an augmented matrix which includes a column of ones. As an additional option we can replace $\phi_j(x)$ with trigonometric terms, e.g.,
\begin{equation}
\begin{split}
        [\hat{Y}_i] = \interval{\underline{w}_0}{\overline{w}_0} &+ \interval{\underline{w}_1}{\overline{w}_1} \sin(\pi x_i) + \interval{\underline{w}_2}{\overline{w}_2} \cos(\pi x_i) + \dots \\
       &+ \interval{\underline{w}_{m-1}}{\overline{w}_{m-1}} \sin(P \pi x_i) + \interval{\underline{w}_m}{\overline{w}_m} \cos(P \pi x_i)
\end{split}
\end{equation}
where $P$ is the order of the trigonometric model and $m = 2P$.

These two extensions show how to use a linear model to characterise an nonlinear relationship between the independent variable and the corresponding conditional mean of $Y$. This generality can also be extended to the imprecise case where input data for the dependent variable has the form of intervals. The next section shows examples of such applications.

\section{Nonlinear Regression as a Two-layer Neural Network}\label{nn_app}

\begin{figure}[ht!]
	\centering
	\begin{subfigure}{0.8\textwidth}
		\includegraphics[width=\linewidth]{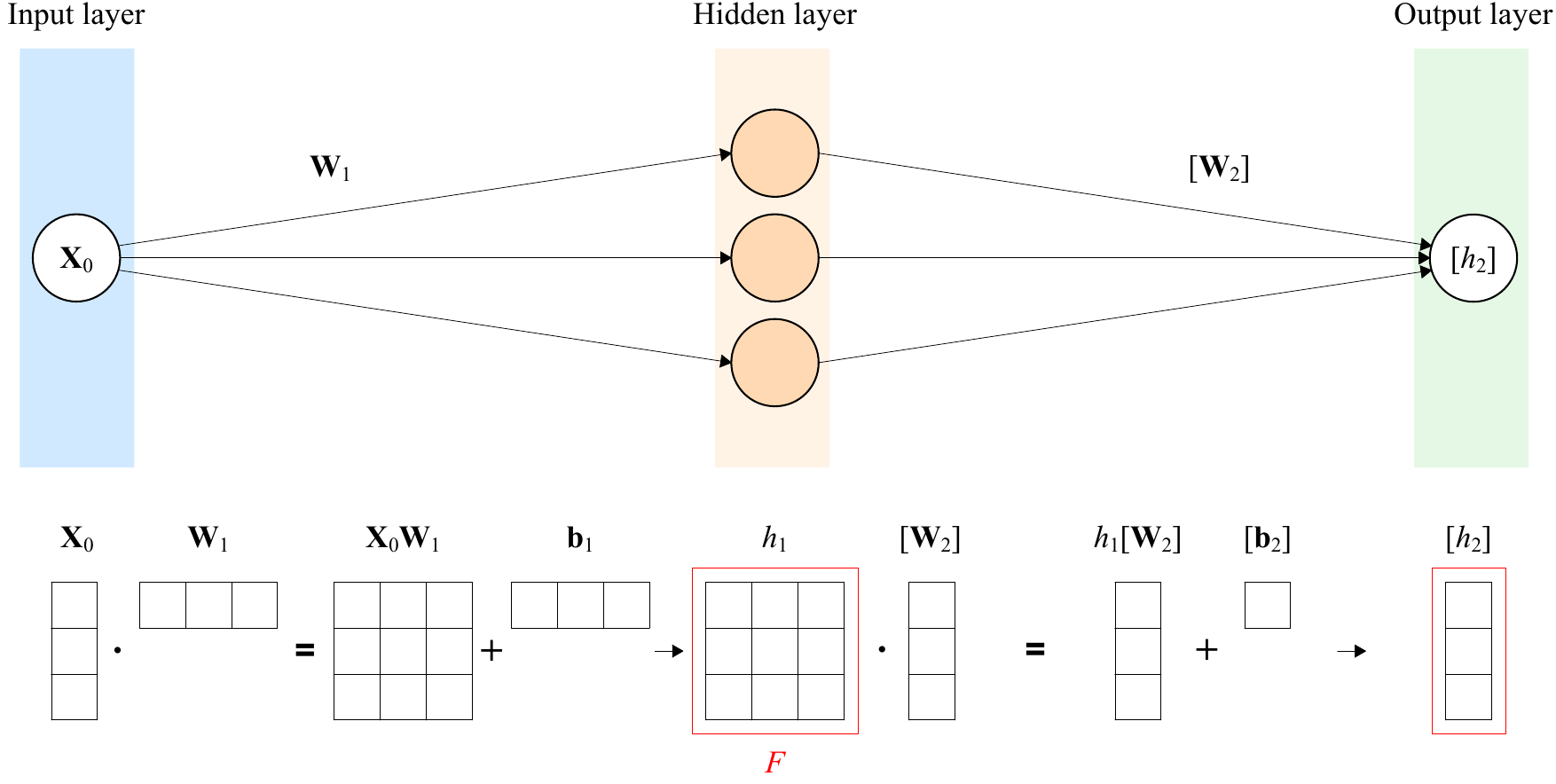}
	\end{subfigure}
	\caption{A feedforward two-layer neural network architecture for a regression setting with interval output. Plus sign denotes `broadcasting' rather than standard matrix addition.
	}
	\label{fig:nn_scheme}
\end{figure}

Exploiting the idea from Section~\ref{forms_of_regression}, we can also use other basis functions which can be more flexible and provide broader tools for constructing imprecise regression models. For instance, let's consider another tunable basis function which is commonly used in artificial neural networks as an activation function $\sigma(x) = 1/(1+e^{-x})$.
We can construct a linear combination of sigmoid functions with some additional internal non-interval parameters $A_j, B_j$ 
\begin{equation}\label{nn_series}
    [h_i] = \sum_{j=1}^m \interval{\underline{w}_j}{\overline{w}_j} \sigma (x_i A_j + B_j) + \interval{\underline{b}_i}{\overline{b}_i},
\end{equation}
or in matrix form
\begin{equation}
    [h] = \bX [\bW]  + [\bb] = 
    \begin{bmatrix}
        \sigma(x_1 A_1 + B_1)  & \dots & \sigma(x_1 A_m + B_m)\\
        \sigma(x_2 A_1 + B_1)  & \dots & \sigma(x_2 A_m + B_m)\\
        \vdots & \ddots & \vdots \\
        \sigma(x_n A_1 + B_1) & \dots & \sigma(x_n A_m + B_m)\\
    \end{bmatrix}
    \begin{bmatrix}
        \interval{\underline{w}_1}{\overline{w}_1} \\
        \interval{\underline{w}_2}{\overline{w}_2} \\
        \vdots \\
        \interval{\underline{w}_m}{\overline{w}_m}
    \end{bmatrix} + [\bb].
\end{equation}
This combination actually represents a feedforward two-layer neural network which consists of layers for hidden and output variables
based on an input layer,
as depicted in Fig.~\ref{fig:nn_scheme}. The coefficients $A_j, B_j$ from (\ref{nn_series}) here are weights and biases of the hidden layer. Using now conventional denotations for the forward propagation we have

\begin{center}
\begin{tabular}{ r c l }
 \textit{Input:} & $\bX_0$ & \\
 \textit{Hidden layer:} & $h_1$ &  $= \; F(\bX_0 \bW_1 + \bb_1)$ \\
 \textit{Output:} & $\intsm{h_2}$ & $= \; h_1 \intsm{\bW_2} + \intsm{\bb_2}$
\end{tabular}
\end{center}
where the sigmoid function $\sigma(x)$ is used for $F$. We use square brackets to emphasize that the hidden layer $h_1$ doesn't contain interval weights and biases unlike the output layer.

The interval \say{linear} model of this two-layer neural network can be efficiently trained to find a set of interval parameters by the two optimization algorithms considered earlier. Some scholars  \cite{Billard_Diday2000,Possibilistic_Approach_2011,Sadeghi2019} 
suggest that a precise result through the interval data can be approximated using central estimator. The  values $\bW_1,\bb_1$ for the hidden layer will be those values that correspond to the central line fitted by  a precise neural network with loss function $\hat{L}$
\begin{equation}\label{nn_precise}
    \hat{h}_2 = F(\bX_0 \bW_1+\bb_1)\, \m \intsm{\bW_2} + \m \intsm{\bb_2} \in \intsm{h_2} ,
\end{equation}
\begin{equation}\label{loss_function2}
    \hat{L} = \dfrac{1}{2} \norm{\hat{h}_2 - \m Y}^2.
\end{equation}
Having defined these $\bW_1,\bb_1$ we can use them to solve for $h_1$ and use the approach previously described for polynomial regression supposing $\bX = h_1$ is constant. Such an approach is not very efficient of course and may be time consuming.  

More efficiently, the training process can be done in one step as in a regular neural network setting. 
Assuming that (\ref{nn_precise}) 
is the midpoint of $[h_2]$, then it is equivalent to the first term in (\ref{mvf_line_shortform}) which represents the central line, hence the resulting interval approximation for the output layer at iteration ($k+1$) is
\begin{equation}
    \intsm{h_2}^{(k+1)}(Y) = \m \intsm{h_2}^{(k+1)} + \mathbb{J}_{h_2} (Y - \m Y),
\end{equation}
\begin{equation}\label{J_h2}
    \mathbb{J}_{h_2} = \dfrac{\dr \intsm{h_2}^{(k+1)} }{\dr Y} = h_1 \dfrac{\dr \intsm{\bW_2}^{(k)}}{\dr Y} + \dfrac{\dr \intsm{\bb_2}^{(k)}}{\dr Y}.
\end{equation}
It should be noted, that equation (\ref{J_h2}) is analogous to (\ref{h_deriv}) except that the latter has a constant matrix $\bX$ instead of layer $h_1$. The Jacobian matrix has only one derivative with respect to $Y$ because we assume that initial guesses for the weights and biases for both layers are certain values, but this could easily be generalised.

\reva{The concept of a network with interval weights and biases in the output layer can be generalized to a multi-layer network as
\begin{equation}
    [h] = [F]^{(\mathcal{L})}( F^{(\mathcal{L}-1)}( F^{(\mathcal{L}-2)}\dots (F^{(l)}\dots(F(\textbf{X}_0))))),
\end{equation}
where $l$ is the layer number for $l=1,\dots,\mathcal{L}$ and the square brackets denote interval values.}

%% file: 3-results.tex
\section{Numerical Examples}\label{Numerical_Examples}


\subsection{Linear regression model}

\reva{In order to illustrate the developed iterative method for computing the expectation band for a linear model with imprecise dependant variables, we use a publicly available red wine quality data set \cite{winedataset}. The data set contains a total of 12 variables, which were recorded for 1,599 observations. This section will illustrate a simple bivariate model and a more complex multivariable model with selected variables to predict wine quality based on chemical properties.}

\reva{The dependent variable (assessed wine quality) has a discrete scale with six levels although we suppose that it is a proxy of a continuous variable. Because the wine quality is given as precise points, we assume that every expert provides an interval estimation of wine quality with width corresponding to the difference between levels so that the \say{true wine quality} $y_{\text{true}}$ is within the respective interval bounds. For example, a reported value of 7 is replaced with the interval [6.5, 7.5].}

\begin{figure}[ht!]
	\centering
	\begin{subfigure}{0.49\textwidth}
	\includegraphics[width=\linewidth]{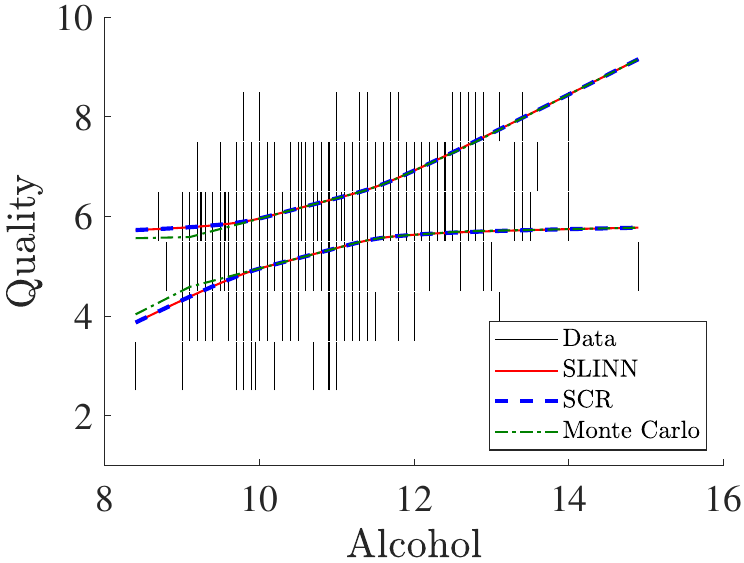}
    \caption{}
	\end{subfigure}
    \begin{subfigure}{0.49\textwidth}
    \includegraphics[width=\linewidth]{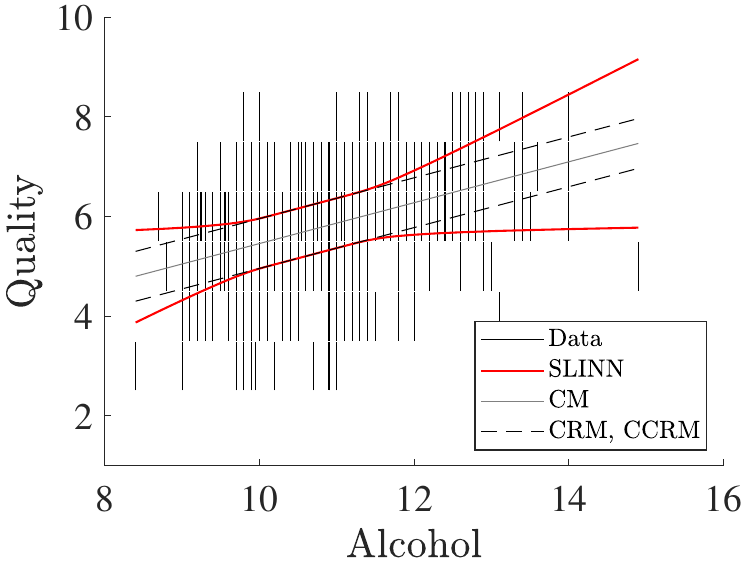}
    \caption{}
	\end{subfigure}
	\caption{\reva{Linear regression example: (a) expectation band obtained from the SLINN (bounded in red lines), from the SCR method (dashed blue lines), Monte Carlo exploration (green dash-dotted lines); (b) the CRM and CCRM methods (dashed black lines) and the CM method produced the solid gray line.}}
	\label{fig:simple-model}
\end{figure}

\textit{Simple model.} \reva{For the simple bivariate model we used \texttt{Alcohol}  amount as the independent variable and computed an imprecise regression model for \texttt{Quality}}.
\reva{The results of applying a single-layer interval neural network with the interval loss function defined in (\ref{loss_function}) is presented in Fig.~\ref{fig:simple-model} and compared to alternative strategies. The network was trained with constant learning rate $\alpha = 5\cdot10^{-4}$, using batch gradient descent optimizer, momentum $\beta=0.9$. The proposed method applied to these interval data yield an imprecise regression which is an expectation band (outlined with red curves).}
\reva{Because this regression model is linear, the single-layer interval neural network and the sharp collection region \cite{SCHOLLMEYER2015} coincide and give the same interval bounds.} The envelope of all  regressions based on any possible configuration of points within the respective data intervals obtained from Monte Carlo simulation with $10^7$ replications is depicted by dash-dotted green lines. \reva{The lower and upper bounds obtained from the center and range method are depicted by black dashed lines. The CM \cite{Billard_Diday2000} model has a slope but does not address the breadth of imprecision in the data.  In contrast, the CRM \cite{Carvalho_2004} model yields two straight lines but seems to understate the uncertainty at extreme $x$-values. The CCRM model gives exactly the same result as the CRM model.} \reva{Table~\ref{table:1} shows computed interval coefficients for the simple model. All the center and range methods are able to predict the lower and upper bound of the interval
value of the dependent variable, but do not give interval coefficients.}
\begin{table}[ht]
\centering
\caption{\reva{A table with interval coefficients for the simple imprecise linear model.}}
\begin{tabular}[t]{lccc}
\hline
 & SLINN & SCR & Monte Carlo\\
\hline
(intercept)   &  [  -2.75,   5.46]  & [-2.75, 5.46]  & [-2.58, 5.29]\\
\texttt{Alcohol}       &  [   0.02,   0.79]  & [ 0.02, 0.79]  & [ 0.03, 0.78]\\
\hline
\end{tabular}
\label{table:1}
\end{table}


\textit{Selected model.} \reva{Based on a correlation analysis we chose a subset of only highly correlated independent variables and performed multivariable linear regression to build an optimal prediction model for red wine quality. Thus, from the initial data set we obtained the selected model which includes 5 chemical properties (\texttt{volatile acidity}, \texttt{citric acid}, \texttt{chlorides}, \texttt{sulphates}, and \texttt{alcohol}). Table~\ref{table:2} shows the obtained interval coefficients for the selected model, and provides a comparison between the proposed method, the sharp collection region and Monte Carlo replications.} 
\begin{table}[ht]
\centering
\caption{\reva{Summary of results from the selected model. The values are interval coefficients from the linear regression.}}
\begin{tabular}[t]{lccc}
\hline
& SLINN & SCR & Monte Carlo\\
\hline
(intercept)         & [   -2.61,    7.48]    & [-2.61 ,7.48 ] & [-1.68, 6.56  ]\\
\texttt{Volatile acidity}    & [  -56.27,   20.55]    & [-56.27,  20.56] & [-46.56, 10.85]\\
\texttt{Citric acid}         & [  -37.93,   40.21]    & [-37.93, 40.21 ] & [-29.85, 32.12 ]\\
\texttt{Chlorides}           & [ -138.01,   73.44]    & [-138.18,  73.53] & [-107.86, 43.21]\\
\texttt{Sulphates}           & [  -18.78,   50.20]    & [-18.78, 50.20 ] & [-14.63, 46.06 ]\\
\texttt{Alcohol}             & [   -1.30,   10.73]    & [ -1.30, 10.73] & [-0.96,10.39 ]\\
\hline
\end{tabular}
\label{table:2}
\end{table}
\reva{The results show that the SLINN can produce similar estimates for interval coefficients as SCR.
The Monte Carlo results suggest these intervals could not be much narrower.}

\subsection{Polynomial regression}

\begin{figure}[ht!]
	\centering
	\begin{subfigure}{0.5\textwidth}
		\includegraphics[width=\linewidth]{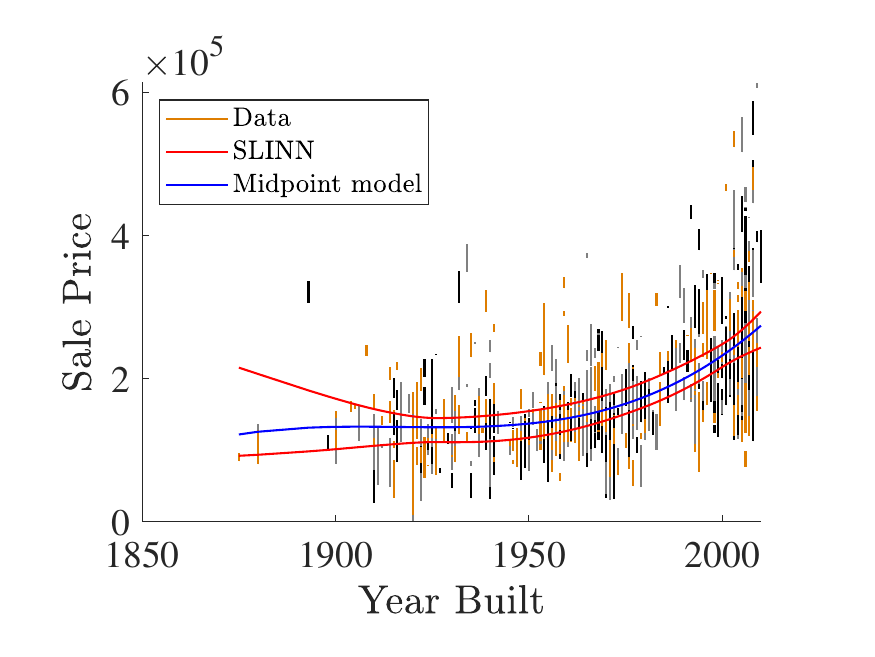}
	\end{subfigure}
	\caption{\reva{Nonlinear regression example. Expectation band (in red) obtained by fitting a 3rd-order polynomial to interval data and a 3rd-order polynomial fitted to midpoints of the data set.}}
	\label{fig:poly}
\end{figure}

\reva{In this section we show the extension of linear imprecise models to capture non-linear relationships described in Section \ref{forms_of_regression}. To illustrate the use of imprecise polynomial regression on interval variables we will use the Ames housing data set \cite{amesdataset}. This data set has 1460 collected samples with multiple features of houses sold during the 1880–2010 years. We fit a 3rd-order polynomial for \texttt{Sale Price} as a function of \texttt{Year Built}. The single-layer network was trained with constant learning rate $\alpha = 10^{-3}$, using batch gradient descent optimizer, momentum $\beta=0.9$ and 3000 epochs. The values of \texttt{Sale Price} are intervalized using the uniformly biased approach so that $[y_i] = \interval{m_i - \Delta}{m_i + \Delta}$, where $m_i = y_{\text{true}} - \Delta  + (1 + b) \Delta$, $b \sim \mathcal{U}(-1,1)$ and $\Delta = \mathcal{N}(15000,10000^2)$ is the interval radius. A plot of the fitted imprecise regression model is presented in Fig.~\ref{fig:poly}.  The expectation band is marked by red solid lines, and the precise regression  model based on midpoints is marked by a blue line. The interval data have variable color (black, gray and brown) for better display.}

\subsection{Neural network numerical examples}
\reva{\textit{Artificial data sets.}} In order to illustrate the developed iterative approach for training multiple-layer neural networks in an imprecise regression setting we first consider two \reva{artificial interval data sets derived from the following analytical functions}: 
\begin{enumerate}
    \item  $y_i = \sin(2\pi x^2) e^{-x} + 1 + \mathcal{N}(0,0.15^2)$, $  \Delta^{\varepsilon}_i \sim |\mathcal{N}(0.1,0.05^2)|$,
    \item  $ y_i = 5 x e^{-3 x} + 0.25 + \mathcal{N}(0,0.02^2)$,  $\Delta^{\varepsilon}_i = 0.2 \exp(-10 (x-0.7)^2)$,
\end{enumerate}
where $\Delta^{\varepsilon}_i$ is the interval radius so that $[y_i] = [y_i-\Delta^{\varepsilon}_i,y_i+\Delta^{\varepsilon}_i]$ and $x_i \sim \mathcal{U}(0,1)$.

\begin{figure}[ht!]
	\centering
	\begin{subfigure}{0.49\textwidth}
		\includegraphics[width=\linewidth]{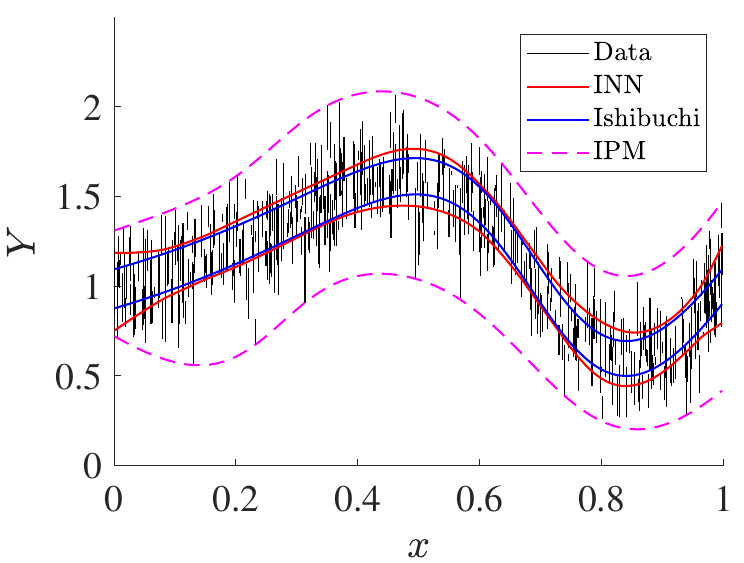}
		\caption{}
	\end{subfigure}
	\begin{subfigure}{0.49\textwidth}
		\includegraphics[width=\linewidth]{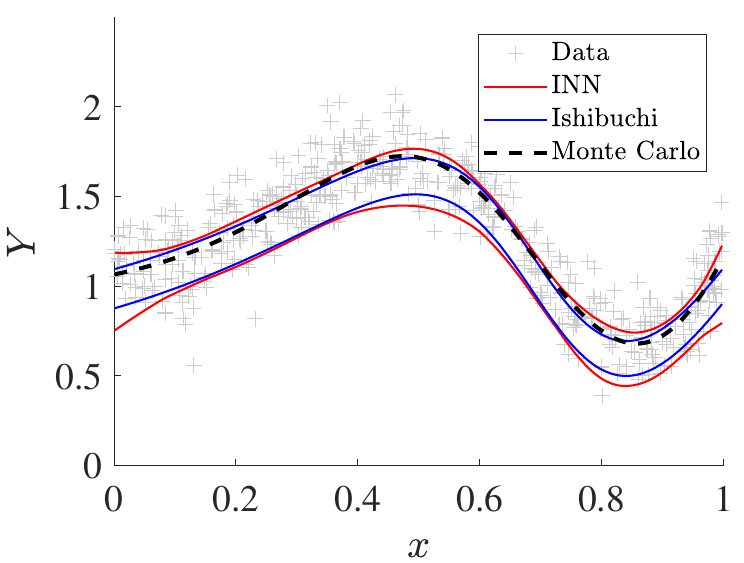}
		\caption{}
	\end{subfigure}
	\caption{\reva{Fitted imprecise nonlinear models: (a) expectation band by the proposed method (red lines), interval output (blue lines) from Ishibuchi's network and bounds obtained by the IPM (magenta dashed lines) (b) output from the precise neural network (dashed black line) trained on data points (gray crosses) taken from the respective intervals.}}
	\label{fig:nn_example1}
\end{figure}

For the test case \#1 we trained 3 different interval neural networks on $N=500$ samples with 1 hidden layer containing 8 nodes. \reva{We use a mini-batch gradient descent optimizer with momentum, a sigmoid activation function and a constant learning rate. The weights were initialised using the uniform initialization $\mathcal{U}(-0.5,0.5)$. Summaries of the hyperparameters used in the numerical experiments are shown in Table~\ref{table:3}.}
\begin{table}[ht]
\centering
\caption{\reva{Summary of the hyperparameters used in the numerical experiments for the simple test functions.}}
\begin{tabular}[t]{lccc}
\hline
Parameter / Configuration & \#1 & \#2 \\
\hline
Hidden layers        & 1 & 2 \\
Number of nodes       & \{8\} & \{16,16\}\\
Activation function           & \texttt{Sigmoid} & \texttt{Tanh}\\
Mini batch size, $M$       & 250 & 200\\
Learning rate, $\alpha$          & 0.005 & 0.005\\
Momentum, $\beta$       & 0.9    & 0.9\\
Number of training epochs           & 6000 & 8000\\
\hline
\end{tabular}
\label{table:3}
\end{table}
\reva{Fig.~\ref{fig:nn_example1}a shows fitted imprecise nonlinear models to interval data set by the proposed method (INN), IPM \cite{Sadeghi2019} and Ishibuchi's method \cite{Ishibuchi_1993}.}
\reva{The expectation band (red lines) produced by the proposed method includes the interval output (blue lines) from the classical interval network which exploits the loss function defined as a sum of the differences between the upper and lower bounds separately. The IPM network produces robust bounds (dashed magenta lines) for all possible values in the data set.}
\reva{Fig.~\ref{fig:nn_example1}b shows a collection of points (gray crosses) taken randomly from the respective intervals in Fig.~\ref{fig:nn_example1}a
with bias toward higher values. The result of using a precise neural network with the same architecture is depicted by the black dashed line. It is clearly visible that there are several regions where the dashed black line is outside of the blue intervals.}

\begin{figure}[ht!]
	\centering
	\begin{subfigure}{0.49\textwidth}
		\includegraphics[width=\linewidth]{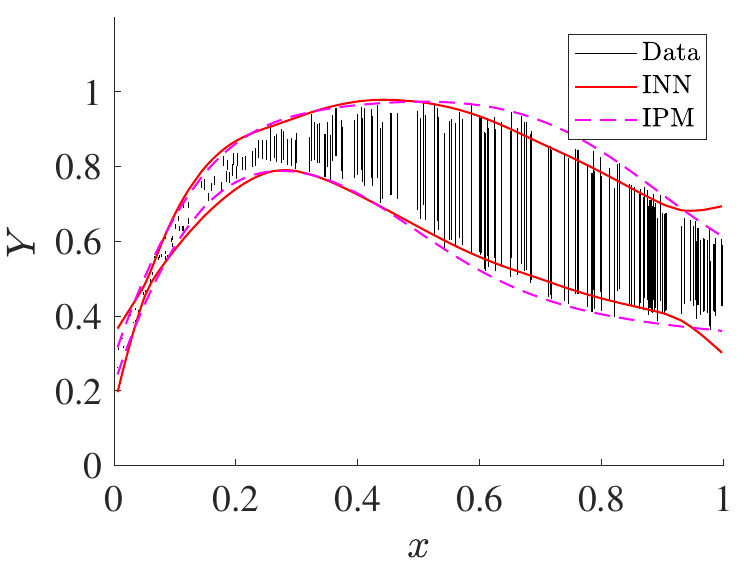}
	\end{subfigure}
	\caption{\reva{Expectation band by the proposed method (red lines) and interval output and bounds obtained by the IPM (magenta dashed lines).}}
	\label{fig:nn_example2}
\end{figure}

\reva{For the test case \#2 we trained our INN and IPM on $N=200$ samples with 2~hidden layers, 16~nodes in each, and a hyperbolic tangent activation function. The hyperparameters are summarised in Table~\ref{table:3}. Fig.~\ref{fig:nn_example2} shows the computed expectation band (red lines) and interval bounds from the interval predictor model (dashed magenta lines).}
\reva{Both the INN and the IPM methods seem to track not just the trend but also the change in imprecision.}

\revb{\textit{Engineering data set.} As a next example we use the concrete compressive strength data set which determines the quality of concrete \cite{YEH1998,CCdataset}. The data set has 1030 collected samples with 9 attributes, where 8 of them are input variables (ingredients and age) and the target variable is concrete compressive strength (\texttt{CC-Strength}). We used all 8 variables to obtain the expectation band for the compressive strength of the concrete as a function of \texttt{Cement}, \texttt{Age}, \texttt{Super Plasticizer}, 
\texttt{Blast Furnace Slag}, \texttt{Fly Ash}, \texttt{Coarse Aggregate}, \texttt{Fine Aggregate} and \texttt{Water}.  An expectation band, rather than a precise regression line, is needed because a rigorous interval approach is often important in civil engineering application and reliability analyses where extreme values are the primary focus of concern.}

\begin{figure}[ht!]
	\centering
	\begin{subfigure}{1.0\textwidth}
		\includegraphics[width=\linewidth]{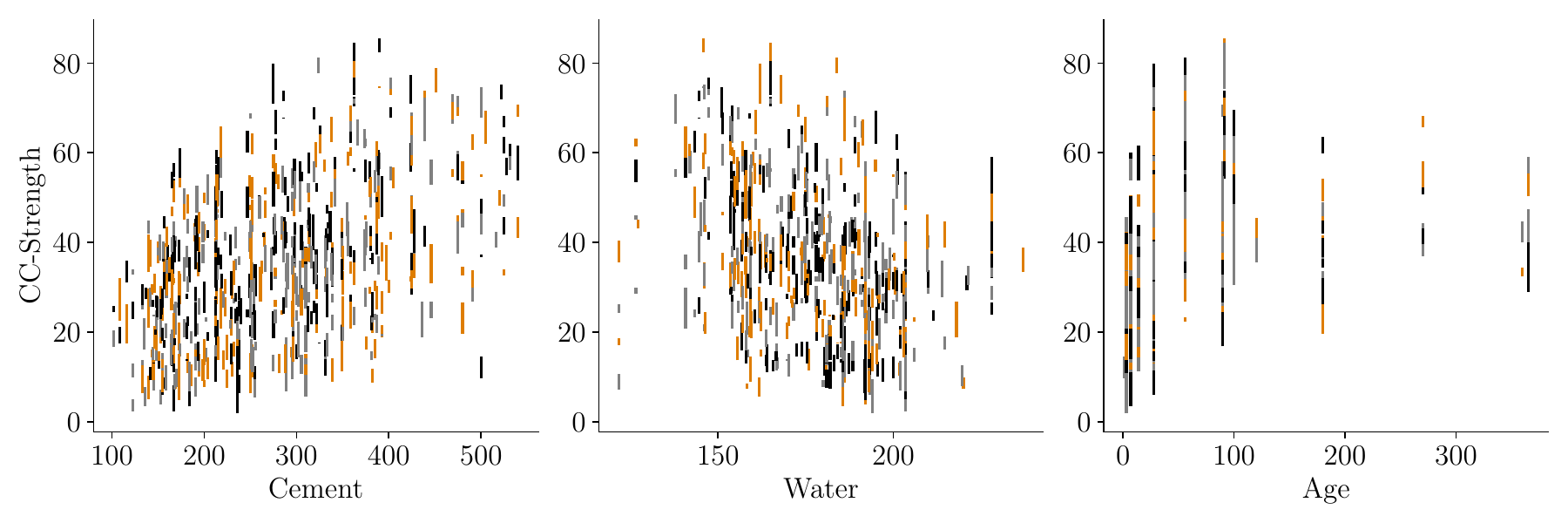}
	\end{subfigure}
	\caption{\revb{Scatter plots for the  intervalized concrete compressive strength as a function of \texttt{Cement}, \texttt{Age} and \texttt{Water}. The interval data have variable color (black, gray and brown) for better display.}}
	\label{fig:CC-strength_data}
\end{figure}

\revb{The values of \texttt{CC-Strength} are intervalized using the uniformly biased approach so that $[y_i] = \interval{m_i - \Delta}{m_i + \Delta}$, where $m_i = y_{\text{true}} - \Delta  + (1 + b) \Delta$, $b \sim \mathcal{U}(-1,1)$ and $\Delta = \mathcal{N}(2,1)$ is the interval radius. In this way we model the interval uncertainty in measurements, which, for example, may arise due to the fact that these measurements are often performed under varying conditions or protocols, or by independent agents, or with different measuring devices. As an illustrative example, three scatter plots between \texttt{CC-Strength} and several features are shown in Fig.~\ref{fig:CC-strength_data}.} 

\reva{We train the proposed algorithm using a train-test split ratio of 0.3, with hyperbolic tangent activation functions on the scaled data set (in the range of 0 to 1) and with decaying learning rate $\alpha = \alpha_0\;\texttt{decay\textunderscore rate}^{-\texttt{step}/\texttt{decay\textunderscore steps}}$, where \texttt{step} is the current epoch. Summaries of the hyperparameters used in the numerical experiment are shown in Table~\ref{table:4}.}
\revb{Summaries of the results from the numerical experiments using training configuratrions \#1-4 are presented in Table~\ref{table:5}. For measuring performance and accuracy of the predictions we used several different interval metrics defined below.}
\revb{The mean minimum and maximum distances $\mathcal{D}_\text{min}, \mathcal{D}_\text{max}$ between interval predictions and targets are defined as
\begin{equation*}
    \mathcal{D}_{\text{min}} = \dfrac{1}{N}\sum_{i=1}^N \inf \{d : d = |a-b|,  a \in [h_i], b \in [Y_i]\}, 
\end{equation*}
\begin{equation*}
\mathcal{D}_{\text{max}} = \dfrac{1}{N}\sum_{i=1}^N \sup \{d : d = |a-b|,  a \in [h_i], b \in [Y_i]\},
\end{equation*}
the mean endpoint distance $\mathcal{D}^*$ is
\begin{equation*}
    \mathcal{D}^* = \dfrac{1}{N}\sum_{i=1}^N \max_i (|\underline{h}_i - \underline{Y}_i|,|\overline{h}_i - \overline{Y}_i|),
\end{equation*}
and the midpoint mean absolute error $\text{MAE}_\text{mid} = \sum_{i=1}^N |\m[h_i] - \m[Y_i]|/N$.}
\begin{table}[ht]
\centering
\caption{\reva{Summary of the hyperparameters used in the numerical experiments for the compressive strength of the concrete data set.}}
\begin{tabular}[t]{lcccc}
\hline
Parameter / Configuration & \#1   & \#2 & \#3 & \#4\\
\hline
Hidden layers                       & 1             & 2             & 1     &   2  \\
Number of nodes                     & \{8\}         & \{8,8\}       & \{24\}&  \{24,24\}   \\
Activation function                 & \texttt{Tanh} & \texttt{Tanh} & \texttt{Tanh}     & \texttt{Tanh}\\
Mini batch size, $M$                & 200           & 200           & 200   &  200     \\
Initial learning rate, $\alpha_0$   & 0.01          & 0.01          & 0.01  &  0.01    \\
Decay rate                          & 0.96          & 0.96          & 0.96  &  0.96    \\
Decay steps                        & 1000          & 1000          & 1000  &  1000    \\
Momentum, $\beta$                   & 0.9           & 0.9           & 0.9   &    0.9  \\
Number of training epochs           & 12000         & 12000         & 12000 &   12000   \\
\hline
\end{tabular}
\label{table:4}
\end{table}
\begin{table}[ht]
\centering
\caption{\reva{Interval metrics from the numerical experiments for the concrete compressive strength data set.}}
\begin{tabular}[t]{lcccccccc}
\hline
 Metric / Config. & \multicolumn{2}{c}{\#1} & \multicolumn{2}{c}{\#2} & \multicolumn{2}{c}{\#3} & \multicolumn{2}{c}{\#4}  \\
\hline
        & Train & Test  & Train & Test & Train & Test & Train & Test\\
\cmidrule(lr){2-3} \cmidrule(lr){4-5} \cmidrule(lr){6-7} \cmidrule(lr){8-9}
$\mathcal{D}_\text{min}$    &1.13 & 1.52& 1.66&2.06 & 0.79& 1.09& 0.34 & 0.58 \\
$\mathcal{D}_\text{max}$    &11.28 & 11.59&9.81 &10.05 & 12.26& 12.84& 12.92 & 13.89 \\
$\mathcal{D}^*$      & 7.21 & 7.61 & 6.05 & 6.34 & 8.14 & 8.82 & 8.79 & 9.86\\
$\text{MAE}_\text{mid}$  & 4.94 & 5.34 & 4.93 & 5.27 & 4.80 & 5.34 & 4.04 & 4.77 \\
\hline
\end{tabular}
\label{table:5}
\end{table}

\revb{The $\text{MAE}_\text{mid}$ metric measures how the centers of the predictions and observed values compare with each other. According to the $\text{MAE}_\text{mid}$, training configuration~\#4 has better predictive capability compared to the other configurations in terms of matching the centers of the observed values and their predictions.
The $\mathcal{D}_\text{min}$ likewise says this configuration has the smallest distances between predictions and observations.
Configuration~\#2 has the highest value of $\mathcal{D}_\text{min}$ which means the fit is poorest among the four configurations, in the sense that the mean distance between observed and predicted intervals is smallest.  Notice, however, this configuration has the lowest value of $\mathcal{D}^*$ which
indicates that the network is producing underestimated interval predictions that do not capture all the uncertainty (that is, both the imprecision and scatter) in the dependant variables.
Although the match between the endpoints of the observed and predicted value is relatively good compared to the other configurations, the 
$\mathcal{D}^*$ metric is sensitive only to the interval endpoints, and not to their centers.
Care must be exercised in interpreting $\mathcal{D}^*$.
}

\begin{figure}[ht!]
	\centering
	\begin{subfigure}{0.49\textwidth}
		\includegraphics[width=\linewidth]{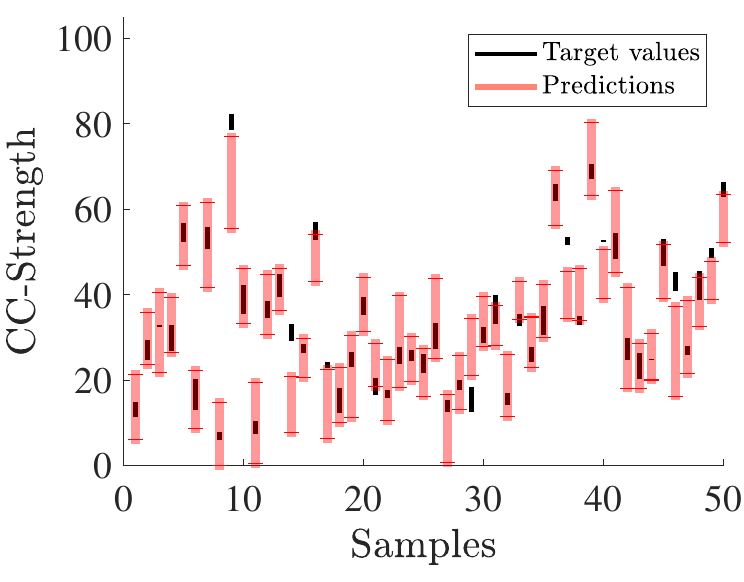}
    \caption{}
	\end{subfigure}
 	\begin{subfigure}{0.49\textwidth}
		\includegraphics[width=\linewidth]{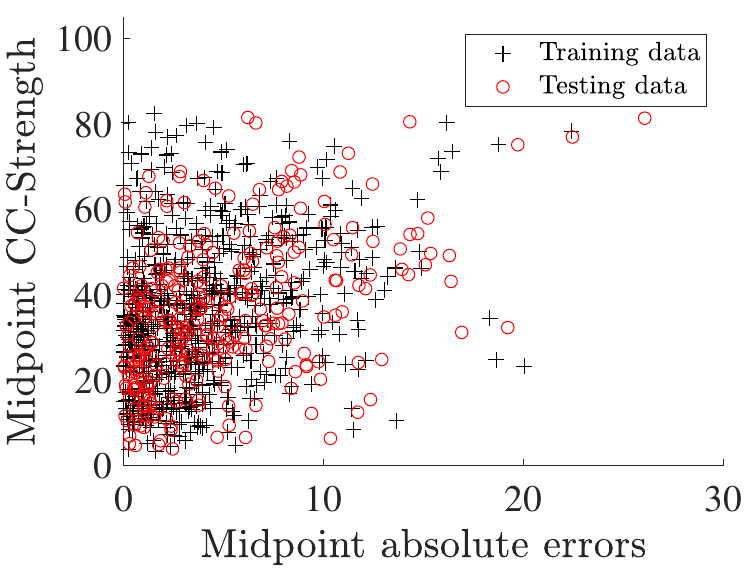}
    \caption{}
	\end{subfigure}
	\caption{\revb{Results from the numerical experiments for the concrete compressive strength data set: (a) 50 randomly selected target variables and their interval predictions from the test data;  (b) midpoint absolute errors for the train and test data.}}
	\label{fig:cc-strength_errors}
\end{figure}

\revb{Fig.~\ref{fig:cc-strength_errors}a shows 50 randomly selected target variables and their interval predictions from the test data for configuration \#4. 
Fig.~\ref{fig:cc-strength_errors}b presents the scatter plot of midpoint absolute errors $|\m[h_i] - \m[Y_i]|$ for the train and test data sets.}
\revb{Both figures show that the network's predictions are in good agreement with target values, and absolute errors for the test and train data have similar trends in scattering.}

%% file: 4-conclusions.tex
\section{Conclusions}\label{Conclusion}

We have presented a new iterative method for fitting an imprecise regression model to data with interval-valued dependant variables. The proposed method is a single-layer interval neural network, a supervised learning algorithm in a regression setting. 
An extended version for nonlinear regressions using a multi-layer interval neural network is also presented which gives stronger tools for modeling uncertainty.
\reva{The main contributions of this paper are that it 
\begin{itemize}
    \item introduces a computationally inexpensive way to compute rigorous bounds on the interval-generalisation of regression analysis to account for epistemic uncertainty in the output (dependent) variables;
    \item introduces the notion of the expectation band as a generalisation of a regression line;    
    \item introduces the use of `neighboring derivatives' to solve certain interval problems involving derivatives in iterated calculations in a  efficient way;
    \item escapes the problem of repeated variables (aka the dependency problem or wrapping effect) even though the loss function is an interval calculation in an iterative process; and
    \item reviews several related ideas in regressions that take account of (epistemic) uncertainty in the $y$-variable.
\end{itemize}}

We have shown several numerical examples in which the interval neural network was trained to quantify uncertainty with interval predictions. The imprecision was modeled non-probabilistically by interval bounds. 
Although interval computations can sometimes be computationally prohibitive, the proposed approach is free of this disadvantage.
Because the interval approach is rigorous, the obtained results will be more reliable than approaches that neglect the imprecision. 

This paper has presented a computationally feasible alternative to both traditional generalised linear regression and regressions based on Bayesian neural networks \cite{A_review_of_UQ2021}, which can be computationally costly. The huge literature from both camps is very diverse, but these applications can be contrasted with proposed method. \reva{The interval neural network does not require probabilistic assumptions, either about imprecision or about the scatter of dependent value, because it models uncertainty with interval bounds.}


The interval bands produced by traditional and Bayesian methods that are created from point estimates of $y$-values arise as confidence or prediction bands (as they are for ordinary least squares regression), not as interval characterisations of measurement imprecision. We defer discussion of the analogous confidence and prediction bands from the proposed approach to a future paper. The traditional and Bayes net methods handle regression when the $y$-values are scattered according to probability distributions, almost always Gaussian distributions. \revb{The proposed approach solves a different problem, which is regression when the uncertainty about the $y$ measurements are characterised by intervals \textit{not} associated with any particular probability distribution. The future paper will discuss the problem of regression for $y$-values that are scattered according to some distribution but are known only through interval measurements.}

\section{Acknowledgements}\label{acks}
We thank Jonathan Sadeghi, Marco de Angelis, Olena Posmitna for their very helpful comments. This work was supported by the UK Engineering and Physical Science Research Council through grant number EP/R006768/1. \reva{The authors would like to thank the anonymous reviewers for their helpful comments and suggestions.}

%% file: 5-appendix.tex
\appendix
\section{Derivatives}\label{app}

Neighbouring derivatives for the gradient descent optimizer in the linear regression setting when $\bW, \bb$ are scalars and $\bX$ is an $n \times 1$ matrix
\begin{align}
    \dfrac{\partial \bW^{(k+1)}}{\partial \bW^{(k)}} &= 1-\alpha \bX^T\bX, \quad    \dfrac{\partial \bW^{(k+1)}}{\partial \bb^{(k)}} = -\alpha \bX^T \mathbf{i}, \quad     \dfrac{\partial \bW^{(k+1)}}{\partial Y} = \alpha \bX^T, \\
    \dfrac{\partial \bb^{(k+1)}}{\partial \bW^{(k)}} &= -\alpha \; \texttt{sum}(\bX), \quad     \dfrac{\partial \bb^{(k+1)}}{\partial \bb^{(k)}} = 1-\alpha \; \texttt{sum}(\mathbf{i}), \quad     \dfrac{\partial \bb^{(k+1)}}{\partial Y} = \alpha.
\end{align}
Operation \texttt{sum}() performs the summation along each column and $\mathbf{i}$ is a column vector of all ones with size $n$. \\

Neighbouring derivatives for the momentum optimizer and a two-layer neural network model
\begin{align}
    \dfrac{\partial \bW^{(k+1)}}{\partial \bW^{(k)}} &= \mathbb{I} - \alpha (1-\beta) h_1^T h_1,  \quad  \dfrac{\partial \bW^{(k+1)}}{\partial \bb^{(k)}} =-\alpha (1-\beta) h_1^T \mathbf{i}, \\
        \dfrac{\partial \bW^{(k+1)}}{\partial \bW^{(k-1)}} &= - \alpha (1-\beta)\beta h_1^T h_1,  \quad  \dfrac{\partial \bW^{(k+1)}}{\partial \bb^{(k-1)}} =-\alpha (1-\beta) \beta h_1^T \mathbf{i}, \\
    \dfrac{\partial \bb^{(k+1)}}{\partial \bW^{(k)}} &= - \alpha (1-\beta) \; \texttt{sum}(h_1), \quad  \dfrac{\partial \bb^{(k+1)}}{\partial \bb^{(k)}} = 1-\alpha (1-\beta) \; \texttt{sum}(\mathbf{i}), \\
    \dfrac{\partial \bb^{(k+1)}}{\partial \bW^{(k-1)}} &= - \alpha (1-\beta)\beta \; \texttt{sum}(h_1), \quad  \dfrac{\partial \bb^{(k+1)}}{\partial \bb^{(k-1)}} = -\alpha (1-\beta)\beta \; \texttt{sum}(\mathbf{i}),
\end{align}
where $\mathbb{I}$ is an identity matrix of size $n$, $h_1$ is a hidden precise layer. The rest of derivatives with respect to $Y$ are 
\begin{equation}
        \dfrac{\partial \bS^{(k)}}{\partial Y} = -h_1^T, \quad \dfrac{\partial \bD^{(k)}}{\partial Y} = -\mathbf{i}^T,
\end{equation}
\begin{align}
    \dfrac{\partial \bW^{(k+1)}}{\partial Y} &= - \alpha (1-\beta) \left[ \dfrac{\partial \bS^{(k)}}{\partial Y} + \beta \eta^{(k)} \right], \\
    \dfrac{\partial \bb^{(k+1)}}{\partial Y} &= - \alpha (1-\beta) \left[ \dfrac{\partial \bD^{(k)}}{\partial Y} + \beta \delta^{(k)} \right],
\end{align}
\begin{equation}
     \eta^{(k)} = \dfrac{\partial \bS^{(k-1)}}{\partial Y} + \beta  \eta^{(k-1)}, \quad \delta^{(k)} = \dfrac{\partial \bD^{(k-1)}}{\partial Y} + \beta  \delta^{(k-1)}.
\end{equation}
